\documentclass[twocolumn,showpacs,preprintnumbers,amsmath,amssymb]{revtex4-1}
\usepackage{graphicx}% Include figure files
\usepackage{dcolumn}% Align table columns on decimal point
\usepackage{bm}% bold math

\begin{document}

\title{Comparative analysis of tunneling magnetoresistance in low-$T_c$ Nb/AlAlOx/Nb and
high-$T_c$ Bi$_{2-y}$Pb$_y$Sr$_2$CaCu$_2$O$_{8+\delta}$ intrinsic
Josephson junctions}

\author{V. M. Krasnov}
\email{Vladimir.Krasnov@fysik.su.se}
\author{H. Motzkau}
\author{T. Golod}
\author{A. Rydh}
\author{S.O. Katterwe}

\affiliation{Department of Physics, Stockholm University, AlbaNova
University Center, SE-10691 Stockholm, Sweden}

\author{A.B. Kulakov}
\affiliation{ Institute of Solid State Physics, Russian Academy of
Sciences, 142432 Chernogolovka, Russia}

\date{\today }

\begin{abstract}

We perform a detailed comparison of magnetotunneling in
conventional low-$T_c$ Nb/AlAlOx/Nb junctions with that in
slightly overdoped Bi$_{2-y}$Pb$_y$Sr$_2$CaCu$_2$O$_{8+\delta}$
[Bi(Pb)-2212] intrinsic Josephson junctions and with microscopic
calculations. It is found that both types of junctions behave in a
qualitatively similar way. Both magnetic field and temperature
suppress superconductivity in the state-conserving manner. This
leads to the characteristic sign-change of tunneling
magnetoresistance from the negative at the sub-gap to the positive
at the sum-gap bias. We derived theoretically and verified
experimentally scaling laws of magnetotunneling characteristics
and employ them for accurate extraction of the upper critical
field $H_{c2}$. For Nb an extended region of surface
superconductivity at $H_{c2}<H<H_{c3}$ is observed. The parameters
of Bi(Pb)-2212 were obtained from self-consistent analysis of
magnetotunneling data at different levels of bias, dissipation
powers and for different mesa sizes, which precludes the influence
of self-heating. It is found that $H_{c2}(0)$ for Bi(Pb)-2212 is
$\simeq 70$ T and decreases significantly at $T\rightarrow T_c$.
The amplitude of sub-gap magnetoresistance is suppressed
exponentially at $T>T_c/2$, but remains negative, although very
small, above $T_c$. This may indicate existence of an extended
fluctuation region, which, however, does not destroy the
general second-order type of the phase transition at $T_c$.


\end{abstract}

\maketitle

\section{Introduction}

Magnetoresistance (MR) is one of the basic tools for analysis of
the electronic structure of metals. MR data contains important
information about {\em bulk} electronic structure of cuprate high
temperature superconductors.
%that is difficult to access by surface probe techniques.
This was demonstrated by recent MR studies
providing compelling evidence for reconstruction of the Fermi
surface in the pseudogap state of underdoped cuprates as a result
of density wave ordering \cite{LeBoeuf2011,Kartsovnik}. Several
experimental techniques revealed the existence of two
distinct energy scales in cuprates: the superconducting %or coherence
gap $\Delta$ and the normal state pseudogap (PG) (for review, see
e.g. Refs.\cite{2gaps,Timusk,NormanRev,STM_review}) with different
behavior with respect to temperature
\cite{2gaps,KrasnovPRL2000,Demsar,Suzuki2223,Bernhard,Shen2007,Raman,SecondOrder},
doping \cite{Doping,Tallon,LeBoeuf2011} and magnetic field
\cite{KrasnovPRL2001,IR_MR_Basov,Lee2006}. However, there is
no consensus on whether they are
competing~\cite{d_dens_H,SecondOrder},
cooperating~\cite{NormanRev}, or representing two manifestations
of the same
phenomenon~\cite{NormanRev,Precursor,STM_review,Ong}.

Analysis of magnetic field effects is particularly useful for
scrutinizing the superconducting origin of the gaps.
Non-superconducting (e.g., structural, antiferromagnetic, charge,
spin, or $d$-density wave) orders are typically insensitive to
achievable fields \cite{d_dens_H}. In this case, magnetic field
may selectively suppress the superconducting gap.
%The main reason is that superconductivity is usually destroyed by
%orbital effects, eg., via acquisition of a large kinetic energy in
%supercurrents flowing around vortices in the mixed state, rather
%than via direct Zeeman-type suppression of the spin-singlet gap by
%magnetic field.
However, discussion of magnetic field effects in cuprates remains
controversial. Conflicting reports exist even on such a basic
parameter as the upper critical field $H_{c2}$. It was reported
that $H_{c2}$ behaves in a conventional manner, i.e., vanishes at
$T_c$ and scales with $T_c$ as a function of doping
\cite{Bouquet,KimHc2,AndoHc2,FluctHc2,TaylorHc2}. But it was also
reported that $H_{c2}$ is $T$-independent and persists well above
$T_c$ \cite{Ong} and increases with underdoping despite reduction
of $T_c$ \cite{KrusinHc2}.

There are many obstacles for deciphering MR data of cuprates such
as: fuzzy superconducting transition due to persistence of the PG;
$d$-wave symmetry of $\Delta$; ill defined quasiparticles (QP) and
strong angular dependence of QP scattering rates \cite{ARPES};
very high anisotropy, which requires accurate control of transport current direction and
sample geometry; existence of an extended fluctuation region above $T_c$
\cite{Varlamov}; possible doping inhomogeneity \cite{STM_review};
and extremely large $H_{c2}\sim 100$ T. One of the important open
question is whether so large magnetic fields simply suppress
superconductivity, or simultaneously induce a competing order, as
may be suggested by observation of charge density
\cite{STM_review} and antiferromagnetic spin order \cite{AF_core}
in vortex cores.

Typically MR involves only QPs at the Fermi level, averaged over
the Brillouin zone, and thus does not provide spectroscopic
information about the QP density of states (DOS) away from the
Fermi surface. A rare exception is the $c$-axis transport in
extremely anisotropic layered cuprates. Single crystals of Bi, Tl
\cite{Kleiner,Warburton} and Hg \cite{Hg} based cuprates represent
natural stacks of atomic scale intrinsic tunnel junctions. The
intrinsic tunnelling spectroscopy (ITS) provides a unique
opportunity to probe directly bulk electronic spectra of cuprates
\cite{Suzuki2223,KrasnovPRL2000,Lee2006,Katterwe2008,SecondOrder}.
Tunneling MR is potentially a very powerful tool for analysis of
superconducting features in electronic spectra. This was
demonstrated in previous studies for cuprates
\cite{Yurgens,KrasnovPRL2001,SuzukiMR,Lee2006,Vedeneev2010} as
well as for conventional low-$T_c$
\cite{Guyon65,Colliner66,Millstein67} and non-cuprate
\cite{Szabo00} high $T_c$ superconductors. Application of magnetic
field leads to appearance of a spatially inhomogeneous mixed
state. This makes analysis of magnetotunneling data non-trivial
\cite{Vekhter,Bulaevskii} and even counterintuitive
\cite{KrPhC2004}. Therefore, a clear understanding of how the
magnetotunneling characteristics of Josephson junctions should
behave is needed for accurate data analysis.

In this work we perform detailed comparison of magnetotunneling in
low-$T_c$ Nb/AlAlOx/Nb, and slightly overdoped
Bi$_{2-y}$Pb$_y$Sr$_2$CaCu$_2$O$_{8+\delta}$ [Bi(Pb)-2212]
intrinsic Josephson junctions with theoretical calculations. Small
sizes of our Bi(Pb)-2212 mesas, which are one-two orders
of magnitude smaller than in previous similar studies
\cite{KrasnovPRL2001,Lee2006,SuzukiMR}, in combination with the
ability to extract information at the sub-gap bias with
low dissipation power, lead to effective obviation of self-heating
\cite{HeatJAP,Heating,SecondOrder}. Both low- and high-$T_c$
junctions show qualitatively similar behavior. Magnetic field and temperature
suppress superconductivity in the state-conserving manner:
enhancement of the sub-gap conductance due to suppression of
$\Delta(T,H)$ is exactly compensated by reduction of the sum-gap
conductance peak. As a result, the MR changes sign from the negative
at the sub-gap voltages to the positive at the sum-gap $eV \gtrsim
2\Delta$. This allows us to trace closing of $\Delta$ at
$T\rightarrow T_c$ with unprecedented clarity. We derive simple
scaling laws for magnetotunneling and employ them for unambiguous
extraction of superconducting parameters such as $\Delta (T,H)$
and $H_{c2}(T)$. The extracted $H_{c2}(T)$ decreases significantly
upon approaching $T_c$. Our data indicates that superconductivity
in slightly overdoped cuprates appears in a conventional manner by
means of the second order phase transition.

The paper is organized as follows. In sec. II we describe the
theoretical formalism, used for microscopic calculation of
current-voltage ($I-V$) characteristics in the mixed state.
Sec. III describes the experimental setup and
studied junctions. Methods used for obviation of self-heating are discussed. In sec. IV we
present main results. It is shown that both Nb and Bi-2212
junctions behave in a qualitatively similar way as a function of
$T$ and $H$.
%In both cases the superconducting gap is suppressed in a state-conserving manner.
In sec. V we analyze scaling laws of different tunneling parameters as a function of $H/H_{c2}$ and
apply them for extraction of $H_{c2}$ in Nb-junctions. This also
provides a clear evidence for persistence of the surface
superconductivity in Nb.
%up to $H_{c3} \simeq 1.69 H_{c2}$.
In sec. VI we apply the same scaling laws for extraction of $H_{c2}(T)$
for Bi(Pb)-2212 and analyze the remaining MR above $T_c$. Finally,
in sec. VII we summarize our conclusions.

\begin{figure*}
\begin{center}
\begin{tabular}{@{} cp{4mm}c @{}}
\includegraphics[width=6.0in]{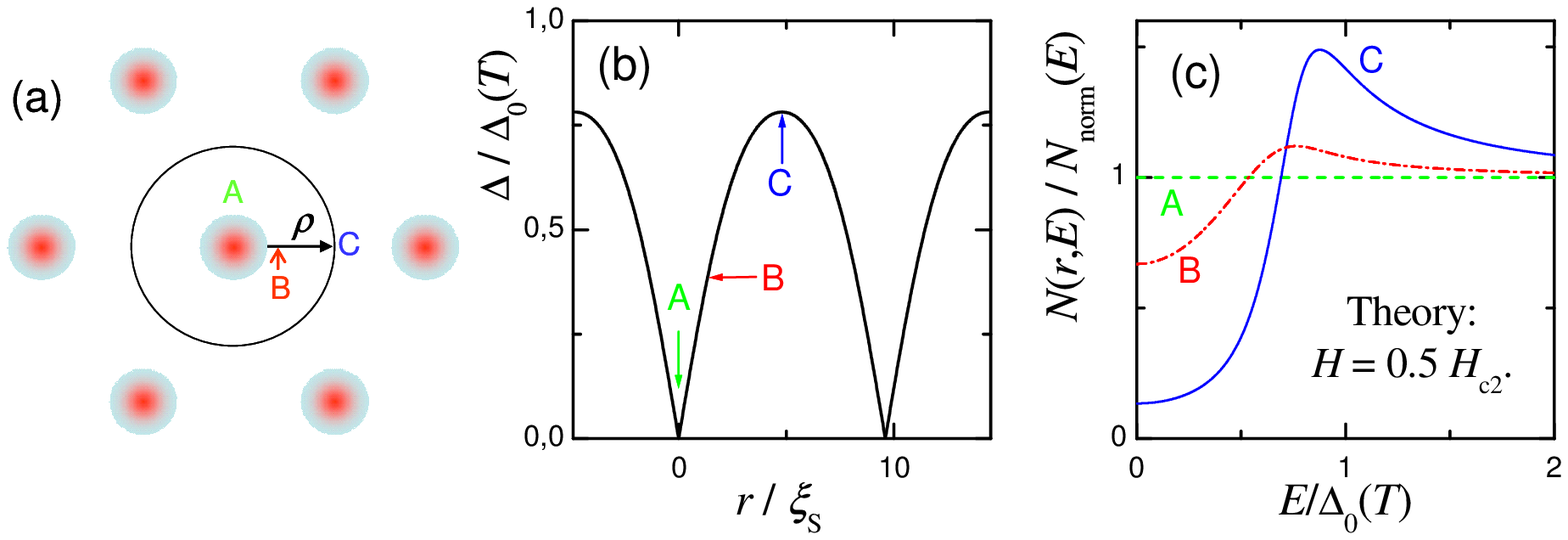}
\end{tabular}
\end{center}
\caption{\label{Fig1_theory} (Color online). (a) A sketch of the
hexagonal Abrikosov vortex lattice and the circular approximation
of a unit cell. (b) Calculated spatial distribution of the order
parameter between two vortices. (c) Local density of states at
three points A, B, C, indicated in (a) and (b). Calculations are made for $H = 0.5~H_{c2}$, $T= 4.7$~K and $T_c=8.8$~K.}
\end{figure*}

\section{Theory}

To calculate spatially non-uniform distribution of the gap and the
DOS in the mixed state, we use the circular cell approximation
\cite{WattsTobinLT,GolubovLT,GolubovPRL}. We approximate the
hexagonal unit cell of the Abrikosov vortex lattice by a circular
cell with a radius corresponding to one flux quantum $\Phi_0$
within the cell, see the sketch in Fig. \ref{Fig1_theory} (a),
\begin{equation}\label{Rs}
\rho=\sqrt{\frac{\Phi_0}{\pi H}}
\end{equation}
and assume a cylindrically symmetric vector potential
\begin{equation}\label{Qr}
Q(r)=1/r - r/\rho^2.
\end{equation}
Spatial distribution of the order parameter $\Delta (r)$ at
different temperatures and magnetic fields are calculated by
solving microscopic Usadel equations: %in Matsubara frequency representation
\begin{eqnarray}\label{Usadel}
\pi T_c \xi_S^2\left[\Theta_n'' +\frac{\Theta_n'}{r}\right] -
\omega_n \sin(\Theta_n)- \\
\nonumber
Q^2(r)\sin(\Theta_n)\cos(\Theta_n) +
\Delta\cos(\Theta_n) = 0,
\end{eqnarray}
together with the self-consistency equation, %for $\Delta$
\begin{equation}\label{SelfConsist}
\Delta \ln(T/T_c) +2\pi T \sum_n\left( \Delta/\omega_n -
\sin(\Theta_n)\right)=0.
\end{equation}
Here $\xi_S=(D/2\pi T_c)^{1/2}$, where $D$ is the QP diffusion
coefficient, $\omega_n=\pi T(2n+1), n=0,1,2,...$ are Matsubara
frequencies, $\cos(\Theta_n)$ and $\sin(\Theta_n)$ are the normal
and the anomalous Green function components, and primes denote
spatial derivation $\partial/\partial r$. Those equations are
subject to boundary conditions at the center of the vortex $r=0$:
$\Delta(0)=\Theta_n(0)=0$, and at the edge of the circular cell
$r=\rho$: $\Delta(\rho)'=\Theta_n(\rho)'=0$. This system of
nonlinear equations was solved numerically, up to a cut-off
frequency $n \simeq 50~T_c/T$, using an iterative procedure. More
details about the formalism and the numerical procedure can be
found in Refs. \cite{GolubovLT,GolubovPRL,GolubovPRB}.

Fig. \ref{Fig1_theory} (b) shows an example of calculated spatial
distribution of $\Delta(r)$ along the line connecting two vortices
for the case of Nb ($T_c=8.8K$) at $T=4.7 K$ and $H/H_{c2} = 0.5$.
The $\Delta(r)$ is normalized by the equilibrium value of the gap
$\Delta_0(T)$ at $H=0$.

To calculate $I-V$ characteristics in the mixed state,
we also need to calculate the spatial variation of the QP DOS
$N(r,E)$ as a function of the QP energy $E$.
This is done by analytic continuation of discrete
$\Theta_n(\omega_n)$ to the continuous energy axis via
substitution, $\omega_n = - i E$, in Eq. (\ref{Usadel}),
where $i$ is the imaginary unit:
\begin{eqnarray}\label{UsadelEn}
\pi T_c \xi_S^2\left[\Theta'' +\frac{\Theta'}{r}\right] + i
E \sin(\Theta)- \\
\nonumber
Q^2(r)\sin(\Theta)\cos(\Theta) + \Delta(r) \cos(\Theta)
= 0.
\end{eqnarray}
The spatially non-uniform DOS is then obtained as
\begin{equation}\label{DOS}
N(r,E)=\Re\left(\cos[\Theta(r,E)]\right).
\end{equation}

Lines in Fig.~\ref{Fig1_theory} (c) show $N(r,E)$,
normalized to the DOS in the normal state, at three points of the
circular cell: A - at the center of the vortex, B - at the
intermediate point where $\Delta(r)$ recovers to half of its
maximum value, see Fig.~\ref{Fig1_theory} (b), and C - at the edge
of the cell. %with maximal $\Delta$.
In the center of the vortex the superconductivity is completely
suppressed, $\Delta =0$, and $N(E) = 1$, as in the
normal state. Away from the vortex the order parameter is partly
restored, but QP spectra are gapless $N(E<\Delta) \neq 0$
\cite{Guyon65,Colliner66,Millstein67}. Due to spatial
inhomogeneity, the maximum in DOS is significantly smeared,
compared to the BCS singularity, and the energy of the maximum is no
longer equal to $\Delta(r)$.

Tunneling $I-V$ characteristics are calculated by integration over
the circular unit cell:
\begin{eqnarray}\label{Eq.Tunn}
%\nonumber
I(V) = \frac{1}{R_n} \int_0^{\rho} \frac{2r}{\rho^2} d r \int_{-\infty}^{+\infty} dE ~~~~~~~~~~\\
\nonumber N(r,E)N(r,E+eV)f(E)\left[1-f(E+eV) \right]. ~~
\end{eqnarray}
Here $R_n$ is the tunneling (normal) resistance and $f$ is the
Fermi distribution function.
%The calculated $I-V$ characteristics as a function of magnetic field and temperature are shown in Fig.\ref{Fig2TheorNb} (a) and(c), respectively.

\subsection{Validity of the model}

The described formalism is valid for arbitrary $T$ and $H$ for
dirty type-II superconductors with $s$-wave symmetry of the order
parameter, which is appropriate for sputtered Nb films.
Calculations presented below are made for BCS parameters with $T_c=8.8$~K, typical for Nb. This allows direct comparison with
experimental data for Nb/AlAlOx/Nb junctions.

In Eq. (\ref{Eq.Tunn}) we
disregard possible misalignment of vortices in the two electrodes.
For intrinsic Josephson junctions, due to very high anisotropy and
weak interlayer coupling, such misalignment may be significant at
low magnetic fields \cite{Vekhter}, but could be neglected for
high fields used in this work.

Certain deviations can be expected for Bi-2212 due to the $d$-wave symmetry of the order
parameter, which reduces the sum-gap singularity and affects the
vortex structure \cite{Ichioka,Volovik,Hussey}. However,
theoretical calculations demonstrated \cite{Ichioka,Vekhter} that
the scaling of DOS characteristics as a
function of $H/H_{c2}$, which will be discussed below, is valid
also for $d$-wave superconductors (see e.g. Figs. 5 from Ref.
\cite{Ichioka}). Therefore, the procedure of extraction of
$H_{c2}$ from such a scaling should be also valid for
intrinsic Josephson junctions.

Coexistence of competing order parameters, associated with the
pseudogap, could also affect the tunneling DOS in cuprates
\cite{Maska,ZhuSDW}. An influence of the PG on the intrinsic
tunneling MR was reported for underdoped Bi-2212
\cite{KrasnovPRL2001,Lee2006,SuzukiMR,KrusinHc2,Bulaevskii}.
Namely, with underdoping the sub-gap MR is rapidly reduced
\cite{KrusinHc2} and suppression of superconductivity by magnetic
field occurs in a seemingly non-state-conserving manner
\cite{SuzukiMR}. This may indicate a gradual recovery of the
competing PG order upon suppression of superconductivity, as
observed in the vortex cores \cite{AF_core,STM_review}. To avoid
possible complications, we restrict our analysis to slightly
overdoped Bi(Pb)-2212, for which there is no significant distortion
by the PG \cite{SecondOrder,Zasad2001}.

\section{Experimental}

We study tunneling magnetorestistance in standard low-$T_c$
Nb/AlAlOx/Nb junctions and in small slightly overdoped Bi(Pb)-2212
mesa structures containing few atomic scale intrinsic Josephson
junctions. Measurements were performed in a gas-flow $^4$He
cryostat in a temperature range down to 1.6\,K and magnetic field
$H$ up to 17\,T. Samples were mounted on a rotatable sample holder
with the alignment accuracy better than 0.02$^{\circ}$. Details of
the measurement setup can be found in
Refs.~\cite{Katterwe2008,Superluminal}.

\subsection{Nb/AlAlOx/Nb junctions}

Nb/AlAlOx/Nb junctions were made by the standard HYPRES trilayer
technology \cite{Hypres} with a critical current density of $1000$
A/cm$^2$. A detailed description of junction parameters can be
found in Ref. \cite{NbAlOxNb}. Junctions consist of two sputtered
Nb thin films with thicknesses 150 and 50 nm for base and counter
electrodes, respectively. Due to different thicknesses, electrodes
have slightly different $T_c$. The critical temperature of the
junction is $\simeq 8.8$ K. The junction barrier is formed by
deposition of a thin Al layer with the thickness $\sim 10$ nm on
top of the base electrode, followed by a subsequent oxidation to
form the AlOx tunnel barrier. During oxidation, only the surface
layer of Al is oxidized, leaving the rest of Al intact. This
results in a proximity effect between the bottom Nb layer and Al
\cite{GolubovPRB}. Tunneling occurs between the proximity-induced
superconducting layer of Al and the top Nb layer.
%The studied temperature range is higher than $T_c$ of the Al layer.
A detailed analysis of the proximity effect in Nb/AlAlOx/Nb
junctions
%with different thickness of Al and at different $T$
can be found in Ref. \cite{GolubovPRB}.

Several junctions with different sizes on the same chip were
studied and showed similar results. As for the case of Bi-2212 mesas \cite{SecondOrder},
with increasing junction area the sum-gap kink in $I-V$ becomes excessively sharp and may even develop a
back-bending as a result of progressive self-heating \cite{Comment2}. Self-heating is
effectively obviated by miniaturization of junctions
\cite{HeatJAP,Heating,SecondOrder}. Therefore, in what follows we
show data only for the smallest junction with sizes $\sim
2.5\times 2.5 ~\mu$m$^2$, which is least affected by self-heating.

\subsection{Bi-2212 intrinsic Josephson junctions}

We study small, micrometer-size mesa structures containing few
atomic scale intrinsic Josephson junctions. The mesas are
fabricated on top of Bi-2212 single crystals using
micro/nano-fabrication techniques. Details of the crystal growth and sample
fabrication can be found in Refs. \cite{Kulakov} and \cite{Submicron}, respectively. Several
contacts on top of the crystals allow us to perform three or
quasi-four probe measurement of the pure $c$-axis transport
\cite{Cascade}. Details of measurements and mesa characterization
can be found elsewhere \cite{Doping,Katterwe2008,SecondOrder}.

We present data for two batches of crystals: lead-doped, slightly
overdoped Bi$_{2-y}$Pb$_y$Sr$_2$CaCu$_2$0$_{8+x}$ with $T_c \simeq 89-93$ K and yttrium-doped, slightly underdoped
Bi$_2$Sr$_2$Ca$_{1-x}$Y$_x$Cu$_2$0$_{8+x}$ (Bi(Y)-2212) with $T_c
\simeq 92$ K. It should be said that the $c$-axis phase coherence
in small mesas is not a good measure of $T_c$. The associated
Josephson coupling energy density is small, and the total
energy is decreasing proportional to the mesa
area. Therefore, in small mesas the phase coherence and the
$c$-axis critical current are suppressed by thermal fluctuations
\cite{Submicron,Anticor2007} at temperatures significantly lower
than that for the in-plane transport. A more detailed discussion
on determination of $T_c$ can be found in Ref.~\cite{SecondOrder}.
Both types of crystals have similar optimal $T_{\mathrm c} \simeq
96$\,K. The most noticeable difference between them is in the
$c$-axis critical current density $J_{\mathrm c}
(4.2\,\mathrm{K})\simeq 10^4$ and $10^3$\,A/cm$^2$ for lead- and
yttrium-doped mesas, respectively. This is due to a rapid increase
of $J_{\mathrm c}$ with over-doping \cite{Doping}.

The $c$-axis transport in Bi-2212 is non-metallic due to
interlayer tunneling mechanism of transport, in combination with
the so-called $c$-axis pseudogap. The latter is much more
pronounced than the PG in the $ab$-plane transport
\cite{Timusk,Bernhard,Dubrovka2010} and exists in a broader
temperature \cite{KrasnovPRL2000,Doping} and doping ranges
\cite{STM_review}. In Ref.~\cite{Katterwe2008} it was shown that
in the normal state $T>T_c$, $c$-axis intrinsic tunneling
characteristics exhibit a trivial thermal-activation behavior,
described by just one constant - the effective barrier height.
This puts a question whether the $c$-axis PG represents the real
two-particle gap in the DOS or is just the single QP tunneling
matrix phenomenon. In Ref. \cite{SecondOrder} more subtle features
in intrinsic tunneling characteristics were found beyond the
thermal-activation background. Those appear in the same
temperature region as in the $ab$-plane transport and were
attributed to the genuine two-particle pseudogap in the QP DOS. A
similar conclusion about the existence of two distinct
pseudogap-like phenomena above $T_c$ was also reached in optical
infrared ellipsometry studies \cite{Bernhard,Dubrovka2010}. The PG
complicates analysis of tunneling magnetoresistance because it
makes the superconducting transition fuzzy. Since the main purpose
of this work is to establish how to extract unambiguous
information out of magnetotunneling data, here we will mostly
concentrate on analysis of overdoped Bi(Pb)-2212 crystals, which
according to previous studies are less affected by the PG and are
most close to the conventional BCS-type superconductivity
\cite{Zasad2001}.

\subsection{Obviation of self-heating in Bi-2212 mesas}

Mesas with different sizes were made on the same single crystal.
As in the case of Nb/AlAlOx/Nb junctions \cite{Comment2}, we observe that $I-V$
characteristics of larger mesas are more distorted by self-heating
at large bias \cite{SecondOrder}. To obviate self-heating, we
perform additional miniaturization of mesa structures
\cite{HeatJAP,Heating,SecondOrder} down to sub-micrometer sizes
using Focused Ion Beam trimming \cite{Submicron}.

\begin{figure*}
\includegraphics[width=7.0 in]{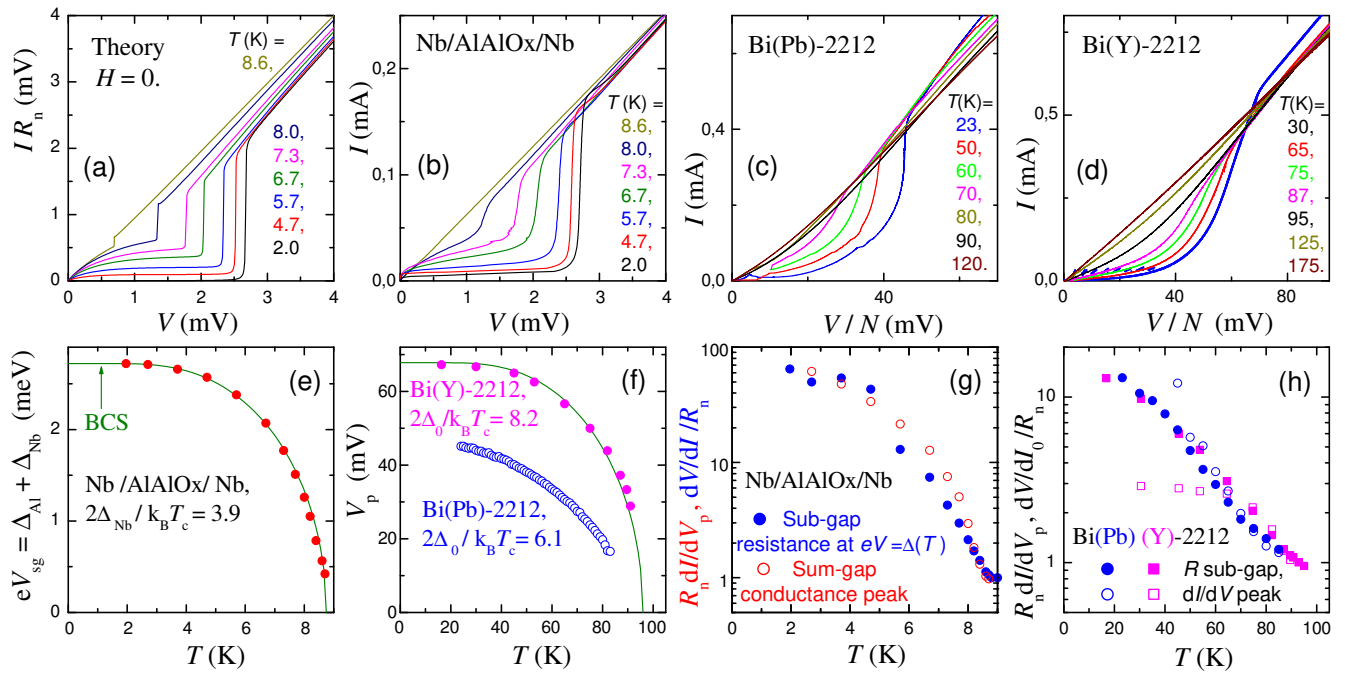}
\caption{\label{Fig2_IVvsT_H0_All} (Color online). Temperature
dependence of tunneling $I-V$ characteristics at zero magnetic field for
(a) calculations for a BCS superconductor (Nb, $T_c=8.8$~K), (b) for the Nb/AlAlOx/Nb junction, (c) for
slightly overdoped Bi(Pb)-2212 mesa and (d) for
slightly underdoped Bi(Y)-2212 mesa. (e) and (f) Measured
$T$-dependence of the sum-gap voltage. Solid lines represent BCS
$T$-dependence. (g) and (h) Comparison of the correlation
between the sub-gap resistance (solid symbols) and the sum-gap conductance (open symbols) for the
same Nb/AlAlOx/Nb junction and Bi-2212 mesas, respectively. }
\end{figure*}

The dissipation powers at the sum-gap knee in $I-V$ (the peak in
$dI/dV$) for the two small mesas shown in Fig.
\ref{Fig2_IVvsT_H0_All} (c) and (d) are $P\simeq 0.13$ and 0.19 mW
for Bi(Pb)-2212 and Bi(Y)-2212 mesas, respectively, at the lowest
temperatures. According to previous reports
\cite{Heating,SecondOrder}, typical thermal resistances of our
micrometer-size mesas lie in the range $\sim 70-30$ K/mW at low
$T$ and decrease to $\sim 10$ K/mW at $T_c$. Thus, self-heating at
the sum-gap peak for those small mesas is manageable and the
effective mesa temperature at the peak remains well below $T_c$.
For the Bi(Y)-2212 mesa this was unambiguously proven by analysis
of the size-dependence of intrinsic tunneling spectra for mesas
with different sizes on the same single crystal
\cite{Heating,SecondOrder}.
To completely exclude possible artifacts of self-heating from the
data analysis, in what follows we define superconducting
parameters from scaling laws valid for tunneling magnetoresistance
at different bias levels, including low, sub-gap bias with
negligible heating.

\section{Results and discussion}

\subsection{Temperature dependence of tunneling characteristics at
zero magnetic field}

Figure \ref{Fig2_IVvsT_H0_All} shows $T$-dependencies of
$I-V$s at $H=0$ for (a) theoretical calculations
for Nb parameters, (b) the Nb/AlAlOx/Nb junction, (c) a slightly
overdoped Bi(Pb)-2212 mesa with the in-plane area $0.9\times 1.3
~\mu$m$^2$ and $T_c\simeq 91$~K and (d) a slightly underdoped
Bi(Y)-2212 mesa with the area $1.8\times 2.0~\mu$m$^2$ and $T_c
\simeq 92$~K. Both mesas contain $N=9$ intrinsic
Josephson junctions, estimated by counting QP-branches in $I-V$ \cite{SecondOrder}.
It is seen that $I-V$s of Bi-2212 mesas are closely resembling
those for conventional superconductor-insulator-superconductor
tunnel junctions. The pronounced sum-gap kink is clearly seen at
low temperatures, followed by an almost $T$-independent
tunneling resistance at higher bias. The sum-gap kink moves to
lower voltage and vanishes in amplitude upon approaching the $T_c$
\cite{KrasnovPRL2000,SecondOrder}.

Experimental characteristics of the Nb/AlAlOx/Nb junction can be
explicitly compared with the corresponding theoretical
simulations. It is seen that there is a good agreement, however, the proximity effect
between Al and Nb leads to some smearing out of the sum-gap kink and enhancement of the sub-gap conductivity.
The most clear proximity induced peculiarity in
$I-V$ of Nb/AlAlOx/Nb is the pronounced dip in $dI/dV$ above the
sum-gap peak (see Fig. \ref{Fig4_dIdVvsH_ALL} b). The dip is caused by the double maxima structure
of the DOS in Al \cite{GolubovPRB}: the lower maximum
corresponds to the proximity induced energy gap in Al,
$\Delta_{Al}$, the upper - to the inherited gap from Nb,
$\Delta_{Nb}$. In this case the sum-gap peak in conductance occurs
at $eV_{sg}=\Delta_{Al}+\Delta_{Nb}$ and the dip at $eV\simeq
2\Delta_{Nb}$. Thus determined gaps are: $\Delta_{Al}=1.22$ meV
and $\Delta_{Nb} = 1.49$ meV, consistent with previous reports \cite{GolubovPRB}.

\begin{figure*}
\includegraphics[width=7.0in]{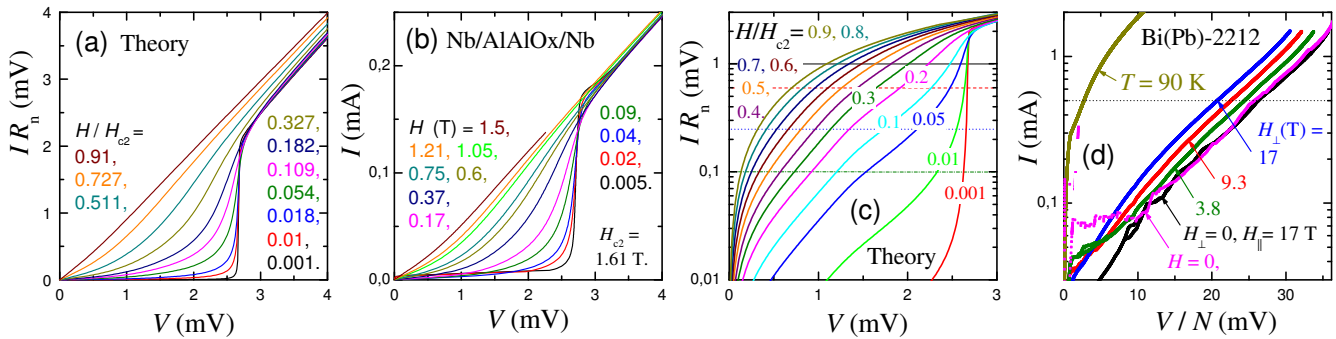}
\caption{\label{Fig3 IV_H_ALL} (Color online). Magnetic field
evolution of $I-V$ characteristics at low $T$ for (a) BCS
calculations at $T\simeq 2$ K and $T_c=8.8$ K, (b) the Nb/AlAlOx/Nb junction at
$T\simeq 2$ K. Semi-logarithmic plots of (c) theoretical $I-V$s at $T\simeq 0.5$ K, and
(d) $I-V$s of a Bi(Pb)-2212 mesa in the
sub-gap region at $T \simeq 2$~K for different $c$-axis field components. Note a
remarkable almost parallel translation of $I-V$s in the
semi-logarithmic scale in (c) and (d).}
\end{figure*}

Due to the $d$-wave symmetry of the order parameter in cuprates,
the sum-gap kinks in Bi-2212 intrinsic tunneling characteristics
are significantly more smeared than in $s$-wave low-$T_c$
junctions. From comparison of Figs. \ref{Fig2_IVvsT_H0_All} (c)
and (d) it is also clear that the sum-gap kink is further loosing
its sharpness with underdoping \cite{Doping}.

Figure \ref{Fig2_IVvsT_H0_All} (e) and (f) represent
$T$-dependencies of the sum-gap voltage (peak in $dI/dV$) at $H=0$
for the Nb/AlAlOx/Nb junction and the Bi-2212 mesas (per intrinsic
Josephson junction), respectively. They were obtained from $I-V$
shown in Figs. \ref{Fig2_IVvsT_H0_All} (b-d). The
bulk gap of Bi(Y)-2212 follows very accurately
the standard BCS dependence. More details on the $T-$dependence of interlayer tunneling
characteristics of our mesas can be found in Refs.
\cite{SecondOrder,Katterwe2008}.

Both Nb and Bi-2212 demonstrate signatures of strong coupling
superconductivity. For Nb the ratio $2\Delta_{Nb}/T_c \simeq 3.9$,
larger than the weak coupling BCS value of 3.5 for $s$-wave
superconductors. For the slightly underdoped Bi(Y)-2212,
$2\Delta_0/T_c \simeq 8.2$. In the slightly overdoped Bi(Pb)-2212, the ratio decreases to 6.1.
This is consistent with previous break junction studies, which indicated that in
overdoped Bi-2212 $2\Delta_0/T_c$ is
approaching the weak coupling BCS value for $d$-wave
superconductors of 4.28 \cite{Zasad2001}.

As seen from Figs.~\ref{Fig2_IVvsT_H0_All} (b-d) in all cases the
increase of $T$ leads to simultaneous increase of the sub-gap
current and decrease of the sharpness of the sum-gap kink. To
demonstrate this reciprocal correlation, in Figs.
\ref{Fig2_IVvsT_H0_All} (g) and (h) we plot the $T-$dependence of
the sub-gap resistance, $dV/dI (V=\Delta/e)$, (solid symbols) and
the sum-gap conductance peak, $dI/dV (V=2\Delta/e)$, (open
symbols) for the same junctions. For Nb/AlAlOx/Nb the correlation is
observed in the whole temperature range. For Bi-2212 the
correlation is holding well until $\sim T_c/2$. The deviations at
lower temperatures are caused by two artifacts affecting the
sum-gap peak height. For Bi(Pb)-2212 the peaks becomes sharper
because of self-heating and for Bi(Y)-2212 it becomes broader
because of minor inhomogeneity of junctions in the mesa
\cite{KrPhysC2002,SecondOrder}.

\subsection{Analysis of tunneling magnetoresistance}

Figs. \ref{Fig3 IV_H_ALL} (a) and (b) show calculated and measured
$I-V$s of Nb/AlAlOx/Nb at $T \simeq 2$~K for different out-of-plane
magnetic fields. Again, a good agreement is seen. With increasing
field, the sub-gap current increases and the sum-gap kink is
rapidly smeared out. The $I-V$ is approaching the ohmic
normal-state at $H\rightarrow H_{c2}$.

From comparison of theoretical curves in Figs.
\ref{Fig2_IVvsT_H0_All} (a) and \ref{Fig3 IV_H_ALL} (a) it is seen
that although both temperature and magnetic field suppress
superconductivity when $T\rightarrow T_c$ and $H\rightarrow
H_{c2}$, there is a difference in how they do that. Namely, the
temperature reduces $\Delta$ but does not affect the shape of the
QP DOS, which remains gapped $N(E<\Delta)=0$ and maintains a sharp
BCS singularity at the gap. Because of that the sum-gap kink
remains sharp even at elevated $T$.
%, although the sub-gap current gradually increases with increasing $T$ because of tunneling of
%thermally activated QP's at $E>\Delta$.
On the other hand, magnetic field first of all smears the gap
singularity in the DOS and the sum-gap kink in $I-V$ and increases
the sub-gap conductance by making the DOS gapless.

Figure \ref{Fig3 IV_H_ALL} (c) shows simulated $I-V$ at different $H$ and at $T=0.5$~K in
the semi-logarithmic scale. It is seen that in this scale the
curves remain almost parallel and move to lower voltage with
increasing $H$ (negative MR).

\begin{figure*}
\includegraphics[width=6.5in]{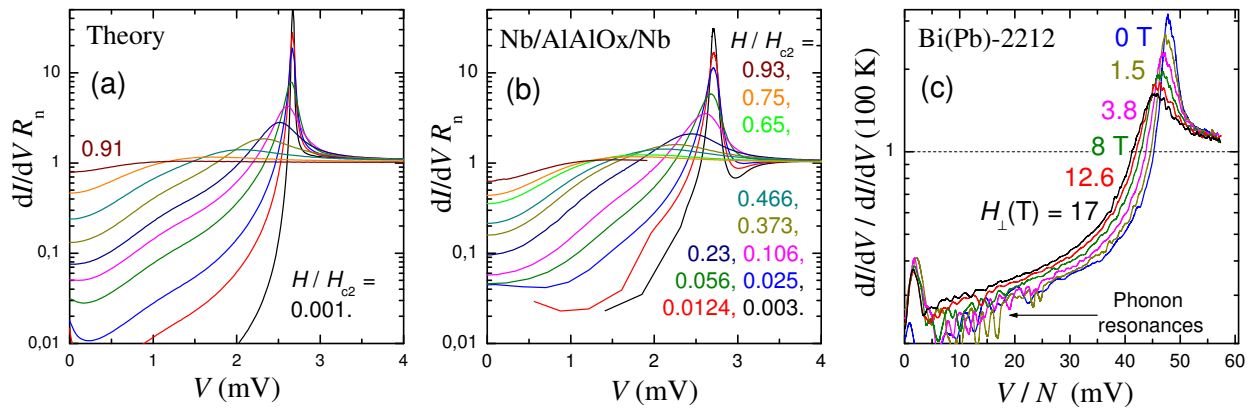}
\caption{\label{Fig4_dIdVvsH_ALL} (Color online). Magnetic field
evolution of $dI/dV(V)$ characteristics at low $T$: (a)
BCS calculations at $T=2$~K, (b) the Nb/AlAlOx/Nb junction at $T=2$~K and (c) a small
Bi(Pb)-2212 mesa (normalized by $dI/dV(T \gtrsim T_c)$) at $T=1.7$~K. All plots
are shown in the semi-logarithmic scale. Low bias structure in (c)
is caused by excitation of phonon-polariton resonances by the ac-Josephson
effect \cite{Polariton}. Due to state
conservation, the magnetoresistance is negative in the sub-gap
region, but positive at and above the sum-gap peak. Note also an almost parallel translation of $\ln dI/dV (V)$ curves with $H$.}
\end{figure*}

Fig. \ref{Fig3 IV_H_ALL} (d) represents the similar
semi-logarithmic plot of $I-V$ curves at different $H$ and at
$T=2$K for a larger Bi(Pb)-2212 mesa on the same crystal as in
Fig. \ref{Fig2_IVvsT_H0_All} (c). Here we changed the $c$-axis
component of the field $H_{\perp}$ from 0 to 17 T by rotating the
crystal with respect to the fixed magnetic field of 17 T. Because
of the extreme anisotropy of Bi-2212, in-plane magnetic field of
17 T does not produce any effect on QP DOS. This is clearly seen
from the two rightmost $I-V$, which were measured at $H=0$
(magenta) and at $H=17$ T strictly parallel to the $ab$-planes
(black). The only minor difference between those curves is caused
by appearance of phonon-polariton resonances \cite{Polariton},
seen as small steps in $I-V$.

Clearly, the general trend of experimental $I-V$ for Bi-2212 is
the same as for numerical simulations and Nb/AlAlOx/Nb: in the
semi-logarithmic scale the curves remain parallel and move to
lower voltages, as the consequence of suppression of the
superconducting gap by field. A similar trend was also observed
for Bi-2212 intrinsic Josephson junctions as a function of $T$
\cite{SecondOrder,Katterwe2008}. For comparison, in Fig. \ref{Fig3
IV_H_ALL} (d) we also show the $I-V$ at $T=90$~K$\simeq T_c$ at
$H=0$. Apparently, the maximum available field of 17 T is
insufficient for complete suppression of superconductivity at
$T=2$ K. The parallel shift of $I-V$ curves implies that the MR,
$V(I,H=0)-V(I,H)$, measured at fixed current and $T$, is
approximately bias-independent. This is important because it
allows a confident estimation of the MR in a broad sub-gap bias
range, including low bias with low dissipation powers, which
precludes distortion by self-heating.

Fig. \ref{Fig4_dIdVvsH_ALL} shows tunneling conductance $dI/dV(V)$
curves in the semi-logarithmic scale at low $T \simeq 2$ K and at
different out-of-plane magnetic fields. Panels (a) and (b) correspond to theoretical
and experimental data for the Nb junction from
Figs. \ref{Fig3 IV_H_ALL} (a) and (b), respectively. At zero field
the main feature of $dI/dV(V)$ curves is the sharp sum-gap peak at $V_{p}=2\Delta/e$, which
reflects the sharp BCS singularity in the QP DOS at $E=\Delta$.
With increasing field, the peak is rapidly smeared
out. Already at $H=0.1~H_{c2}$ the height of the peak is suppressed by
an order of magnitude. Simultaneously the sub-gap conductance at
$V<V_{p}$ is growing with field. The excess QP current is flowing
in gapless vortex cores, see Fig. \ref{Fig1_theory} (a). It scales
with the relative core area in the unit cell and, therefore,
increases approximately linearly towards the normal conductance at
$H \rightarrow H_{c2}$.

In Fig. \ref{Fig4_dIdVvsH_ALL} (c) we show $dI/dV(V)$
characteristics in the semi-logarithmic scale for different
$H_{\perp}$ at $T=1.7$ K, for another mesa $1\times 1.9~\mu$m$^2$ on the same Bi(Pb)-2212 single crystal.
To simplify the analysis, the curves are normalized by the normal state curve at
$T=100$ K, so that the normal state is simply represented by
the dashed horizontal line. It is seen that
$dI/dV(V)$ curves shift as a whole to lower voltages with
increasing field. Simultaneously the peak looses the height while
the sub-gap conductance vice-versa increases with increasing
field. Such the behavior is almost identical to that for the
Nb/AlAlOx/Nb junction and calculations, shown in Figs.
\ref{Fig4_dIdVvsH_ALL} (a) and (b).

\subsection{State conservation}

According to theory, suppression of superconductivity both by
temperature and magnetic field occurs in the state-conserving
manner. Even though superconductor-insulator-superconductor
tunneling is not probing explicitly the single QP DOS, but rather
the convolution of two DOS from the two electrodes,
Eq.~(\ref{Eq.Tunn}), it is still possible to judge about the state
conservation by a simple integration of $dI/dV$ curves:
\begin{equation}\label{StateConserv}
\int_0^{\infty}{\left(R_n\frac{dI}{dV}-1\right)dV}=0.
\end{equation}
The state conservation is the reason for $T$- and $H$-independence of the
large bias tunneling resistance $R_n$, as seen from Figs.
\ref{Fig2_IVvsT_H0_All} (a) and \ref{Fig3 IV_H_ALL} (a).

From Figs.~\ref{Fig2_IVvsT_H0_All} (b-d) it is clear that both in
low-$T_c$ and high-$T_c$ junctions the Ohmic tunneling resistance
$R_n$ above the sum-gap kink is
remaining almost $T$-independent. According to Eq.
(\ref{StateConserv}), this automatically implies that closing of
the superconducting gap by temperature occurs in the
state-conserving matter.

State conservation implies that the tunneling resistance is
enhanced in the sub-gap region in the same manner as the
conductance is enhanced at the sum-gap peak. This is demonstrated
explicitly, in Figs.~\ref{Fig2_IVvsT_H0_All} (g) and (f) for
Nb/AlAlOx/Nb and Bi-2212 junctions, respectively. It is seen that
the sub-gap resistance $dV/dI$ in the middle of the sub-gap region
$eV=\Delta(T)$ and the sum-gap conductance peak grow in a similar
manner with decreasing $T$. Such a scaling is an instructive way
for examining the state conservation in Bi-2212 because it does
not depend on a small $R_n(T)$
dependence~\cite{KrasnovPRL2000,Katterwe2008}, which is likely due
to a minor in-plane (coherent) contribution to the interlayer
transport \cite{Giura}.

\subsection{The sign change of tunneling magnetoresistance}

The tunneling MR in superconducting tunnel junctions and the
$c$-axis MR in layered cuprates is often described as negative.
That is, the resistance decreases with increasing field. This is
the consequence of appearance of the gapless state in magnetic
field. The corresponding increase of the sub-gap DOS leads to
approximately linear increase of the low bias tunneling
conductance with field \cite{Bulaevskii}. This is clearly seen in
Fig. \ref{Fig4_dIdVvsH_ALL}. However it is also seen that at the
sum-gap peak the situation is reversed: the tunneling conductance
is decreasing with increasing field, i.e., the tunneling MR at
$V\gtrsim V_{p}$ is positive. Thus, the tunneling MR is changing
sign from the negative at the sub-gap voltage to the positive close and
above the sum-gap voltage. Again, the behavior is similar for both
Nb/AlAlOx/Nb and Bi(Pb)-2212 junctions.

The sign change of the tunneling MR is a direct consequence of state conservation. The
missing area of the sub-gap conductance with respect to normal
conductance $1/R_n$ is equal to the excess area of the
sum-gap peak. Therefore, the positive MR at large bias is directly
connected with the negative MR at low bias. The discussed
sign-change of the MR is very characteristic and can be used for
unambiguous discrimination of the superconducting gap from
non-pairing effects in the tunneling DOS, such as peculiarities of
one-QP band structure, or thermal activation enhancement of the
tunneling matrix elements for interlayer hopping
\cite{Katterwe2008}.

\begin{figure*}
\includegraphics[width=6.0in]{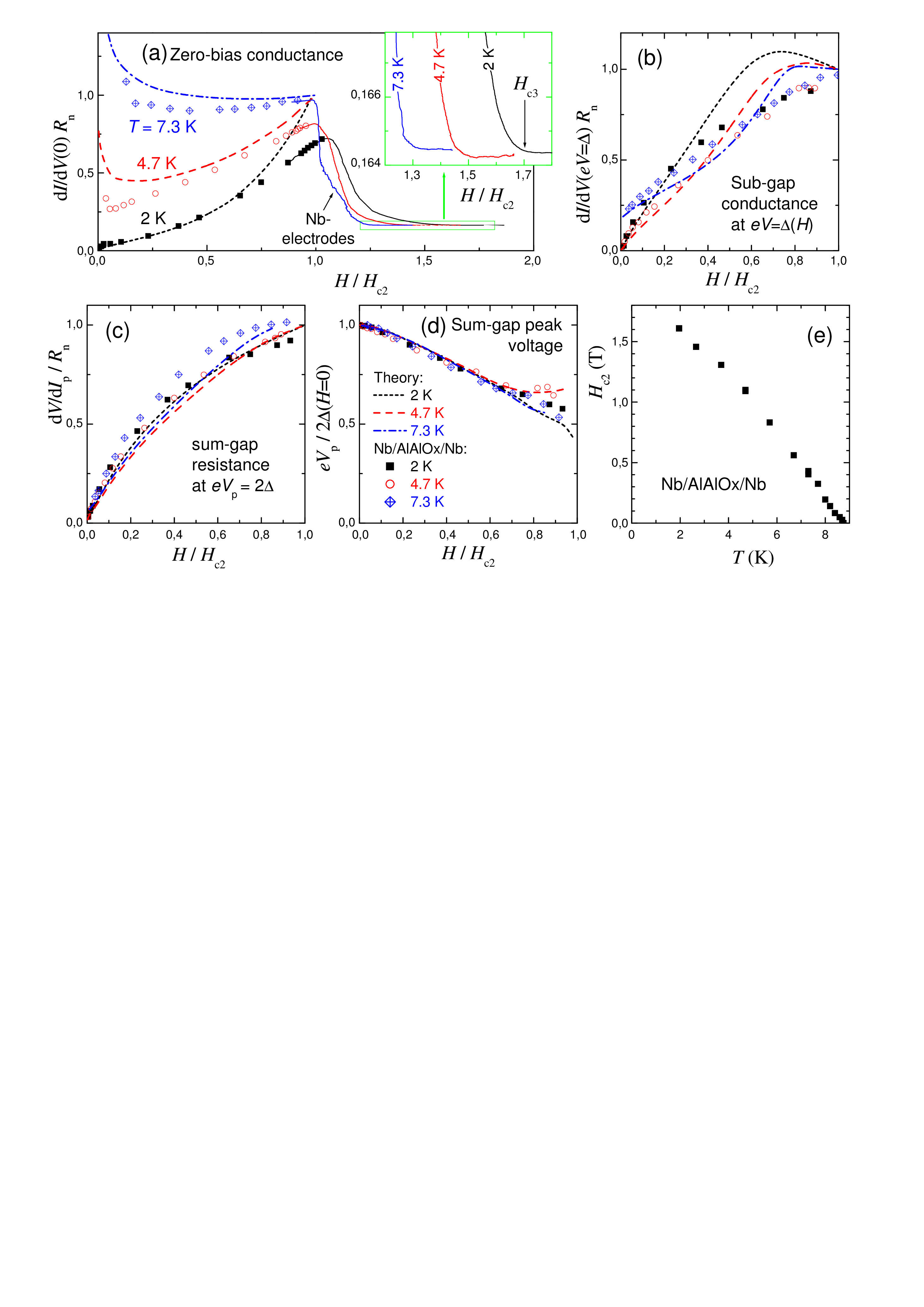}
\caption{\label{Fig5_NbBCSSum} (Color online). Analysis of
scaling of magnetotunneling characteristics at different $T$ as a
function of $H/H_{c2}$ in theory (thick lines) and for the Nb
junction (symbols). (a) Zero bias
conductance. Note that there is no scaling of $dI/dV(0)$ for
different $T$ due to progressive development of the zero-bias
logarithmic singularity with increasing $T$. Thin solid lines and
the inset show evolution of the conductance of Nb-electrodes at
$H>H_{c2}$. It is seen that the surface superconductivity in
electrodes survives up to $H_{c3}\simeq 1.69 H_{c2}$. (b) Scaling
of the sub-gap conductance at finite bias $V=\Delta/e$. (c)
Scaling of the sum-gap resistance. (d) Scaling of the sum-gap peak
voltage. (e) The upper critical field for Nb, extracted from the
scaling.}
\end{figure*}

\section{Scaling laws of tunneling magnetoresistance}

The main purpose of this work is to determine how to extract
useful information from tunneling MR. To
understand this we first consider the behavior of magnetotunneling
for conventional BCS superconductors. For this we analyze
numerically calculated and experimental characteristics for Nb junctions
at different $T$ and $H$.

\subsection{Zero-bias magnetoresistance}

Transport measurements are
usually performed by applying a small ac-current, i.e., probe
zero-bias MR. Thick lines in Fig.~\ref{Fig5_NbBCSSum}
(a) show theoretical values of the zero bias conductance
$dI/dV(0)$, normalized by $R_n$, as a function of $H/H_{c2}$ for
$T = 2$, 4.7 and 7.3 K. It is seen that at low $T=2$~K, the zero
bias conductance is gradually increasing from almost zero at $H=0$, to
normal conductance at $H_{c2}(T)$ (see Fig. \ref{Fig4_dIdVvsH_ALL} (a)).

Temperature dependence of $dI/dV(0)$ at $H=0$ can
be seen from $I-V$ curves in Fig. \ref{Fig2_IVvsT_H0_All} (b). The
increase of $T$ leads to the decrease of $\Delta$. Both factors
result in the increased number of excited QPs above the gap,
which initially leads to a rapid filling-in of the zero-bias dip
in conductance and then to development of a maximum at $V=0$. The
latter represents a zero-bias logarithmic singularity
\cite{Larkin}. It occurs because at
elevated $T$ there is a substantial amount of thermally excited
QPs just above the gap. At $V=0$ the partly filled gap
singularities in the two electrodes are co-aligned, causing a
large current flow from one electrode to another, which is exactly
compensated by the counterflow from the second electrode. However,
exact cancellation is lifted at an arbitrary small voltage across
the junction, leading to a sharp maximum in $dI/dV$. The zero-bias
logarithmic singularity leads to an overshooting of $dI/dV(0)$ at
$H=0$ over the normal conductance at $T=7.3$~K in Fig.
\ref{Fig5_NbBCSSum} (a).

The zero-bias logarithmic singularity makes the
behavior of the zero-bias conductance non-trivial. In general,
there is no scaling with $H/H_{c2}$ for $dI/dV(0)$ at different
$T$, as seen from Fig. \ref{Fig5_NbBCSSum} (a). To avoid
complications caused by the zero-bias singularity we look at the
behavior of the conductance at finite bias.

\subsection{Scaling of the sub-gap conductance and the sum-gap resistance}

Fig. \ref{Fig5_NbBCSSum} (b) shows field dependence of
the sub-gap conductance at the middle point $eV=\Delta$ for the
same temperatures as in (a). It is seen that unlike $dI/dV(0)$,
the sub-gap conductance  is showing a fairly universal
linear scaling as a function of $H/H_{c2}$ for different $T$.

As follows from Figs. \ref{Fig2_IVvsT_H0_All} (g) and (h), state
conservation implies that the deficit of the sub-gap conductance
is directly connected to the excess of the sum-gap peak, i.e., the
deficit of the sum-gap resistance. Fig. \ref{Fig5_NbBCSSum} (c)
shows magnetic field dependence of the sum-gap resistance. It is
showing an almost universal, slightly non-linear scaling as a
function of $H/H_{c2}$ in the wide $T$-range.

Fig. \ref{Fig5_NbBCSSum} (d) shows magnetic field dependence of
the sum-gap peak voltage $V_p$ normalized by that at zero field.
It exhibits a very simple universal linear scaling as a function
of $H/H_{c2}$ in the whole $T-$range. Interestingly, the peak
voltage does not go to zero at $H=H_{c2}$, but rather stops
half-way at $eV \simeq \Delta (H=0)$. This is due to an interplay
between the reduction of $\Delta(H)$, see
Fig.~\ref{Fig1_theory} (b), which moves the peak down, and
simultaneous strong smearing of the maximum in the QP DOS, see
Fig.~\ref{Fig1_theory} (c), which moves the peak up in voltage. At
$H=H_{c2}$, $\Delta$ does vanish, but voltages/energies of very
broad maxima in $dI/dV$ or spatially averaged DOS do not vanish
\cite{KrPhC2004}, as can be seen from Fig. \ref{Fig4_dIdVvsH_ALL}
(b).

\subsection{Extraction of $H_{c2}$ for Nb}

The universal scaling of the sum-gap peak voltage
$V_p(H/H_{c2})$ can be used for extraction of $H_{c2}$ from magnetotunneling data. Symbols in
Fig.~\ref{Fig5_NbBCSSum} (d) show the results of fitting for the
Nb/AlAlOx/Nb junction, using $H_{c2}(T)$ as the only adjustable
parameter. It is seen that the agreement with theoretical
calculations is excellent. $T$-dependence of the obtained upper
critical field is shown in Fig.~\ref{Fig5_NbBCSSum} (e). Due to
the proximity effect in Al it is more linear at low $T$ than that
for pure Nb.
%and is close to $H_{c2}(T)/H_{c2}(0) \simeq 1- (T/T_c)^{1.5}$.
A similar $H_{c2}^{\perp}(T)$ was also reported for
proximity coupled Nb/Cu multilayers \cite{NbCu}.

Using thus extracted $H_{c2}$, we check the scaling of
other experimental parameters such as the zero bias and the
sub-gap conductances and the sum-gap resistance, shown by symbols
in Figs. \ref{Fig5_NbBCSSum} (a), (b) and (c). In
all cases the agreement with theory is very
good, confirming the correctness of determination of $H_{c2}$.

\begin{figure*}[t]
\begin{center}
\begin{tabular}{@{} cp{4mm}c @{}}
\includegraphics[width=6.5in]{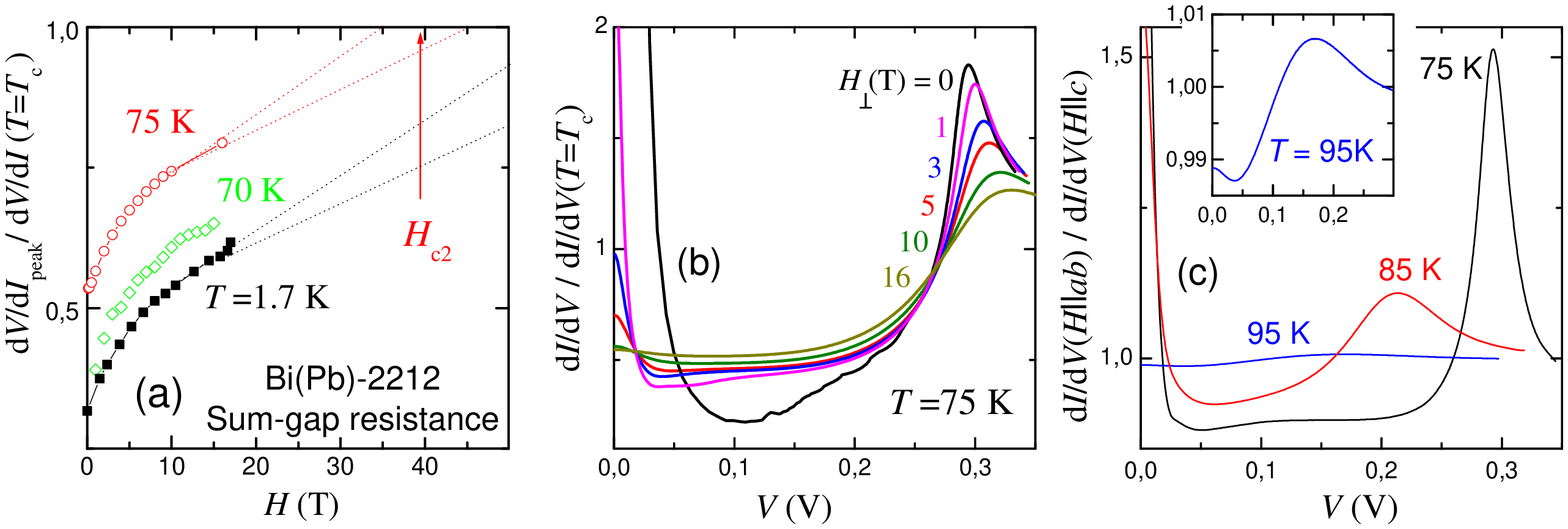}
\end{tabular}
\end{center}
\caption{\label{Fig6_IVH_AR} (Color online). (a) Field-dependence of
the sum-gap resistance in Bi(Pb)-2212 at different $T$ (normalized to the normal state value). Dotted lines indicate
extrapolations to $H=H_{c2}$. (b) Magnetic field
evolution of $dI/dV(V)$ curves at $T=75$ K (normalized by $dI/dV(V)$
at $T\simeq T_c$). (c) Relative change of $dI/dV$ upon rotation of the crystal in
field $H=10$ T from the in-plane to the $c$-axis direction. Inset shows
enhanced view of the curve at $T=95$ K. }
\end{figure*}

\subsection{Surface superconductivity above $H_{c2}$ in Nb}

The only case where there is a certain disagreement between theory and experiment
is $dI/dV(0)$ at high fields in Fig.~\ref{Fig5_NbBCSSum} (a): experimental data at
$T=2$~K and 4.7~K is clearly not reaching the normal conductance
at $H=H_{c2}$. At higher fields it is not possible to measure
junction characteristics because the Nb electrodes are no longer
capable to carry the supercurrent and turn into the resistive
state. Since the electrode resistance is several times larger than
that of the junction, the measured resistance in this case is
mostly given by the longitudinal resistances of the Nb electrodes.
This leads to the drastic decrease in the measured zero-bias
conductance, as shown by thin solid lines in Fig.
\ref{Fig5_NbBCSSum} (a). From this data it is obvious that some
residual superconductivity in the Nb electrodes is remaining up to
fields significantly larger than $H_{c2}$.

The inset in Fig.~\ref{Fig5_NbBCSSum} (a) shows the detailed view
of the onset of the superconducting transition in Nb electrodes.
It is seen that at low $T=2$~K superconductivity in Nb survives up
to almost exactly $1.69 H_{c2}$, which is the expected value of
the third critical field $H_{c3}$. At $H_{c2}<H<H_{c3}$
superconductivity exists only in surface layers. A similar
behavior has been reported in clean Nb
\cite{Hc3NbPark2003,Hc3NbKotzer2004,Hc3NbDas2008} as well as in
MgB$_2$ \cite{Hc3MbB2}. From the inset in Fig. \ref{Fig5_NbBCSSum}
(a) it is seen that the ratio $H_{c3}/H_{c2}$ is decreasing with
increasing temperature. A similar behavior was reported for clean
Nb and discussed in terms of a tricritical point
\cite{Hc3NbPark2003}. However, it should be noted that our Nb
films are in the dirty limit and are affected by the proximity
effect with Al. The columnar structure of sputtered Nb films with a
large effective surface-to-volume ratio and columns orientation
perpendicular to the film may also enhance the role of surface
superconductivity in the out-of-plane magnetic field.

\section{Analysis of intrinsic magnetotunneling in Bi-2212}

Observation of a fairly conventional behavior of tunneling MR in
overdoped Bi(Pb)-2212 encourage us to employ the derived scaling laws for
extraction of the $T$-dependent upper critical field,
which remains a controversial issue for cuprates, as mentioned in the
introduction. The obtained scaling of the sum-gap peak voltage $V_p(H/H_{c2})$
is valid not only for tunnel junctions made of $s$-wave
superconductors. A similar scaling was also reported for
$\Delta(H/H_{c2})$ in the mixed state of $d$-wave superconductors
\cite{Ichioka}. In the remaining part of this work we apply the
scaling rules for Bi(Pb)-2212 mesas in order to understand how
normal or abnormal the behavior of intrinsic tunneling
magnetoresistance is and in an attempt to estimate the upper
critical field.

\begin{figure*}[t]
\includegraphics[width=7.0in]{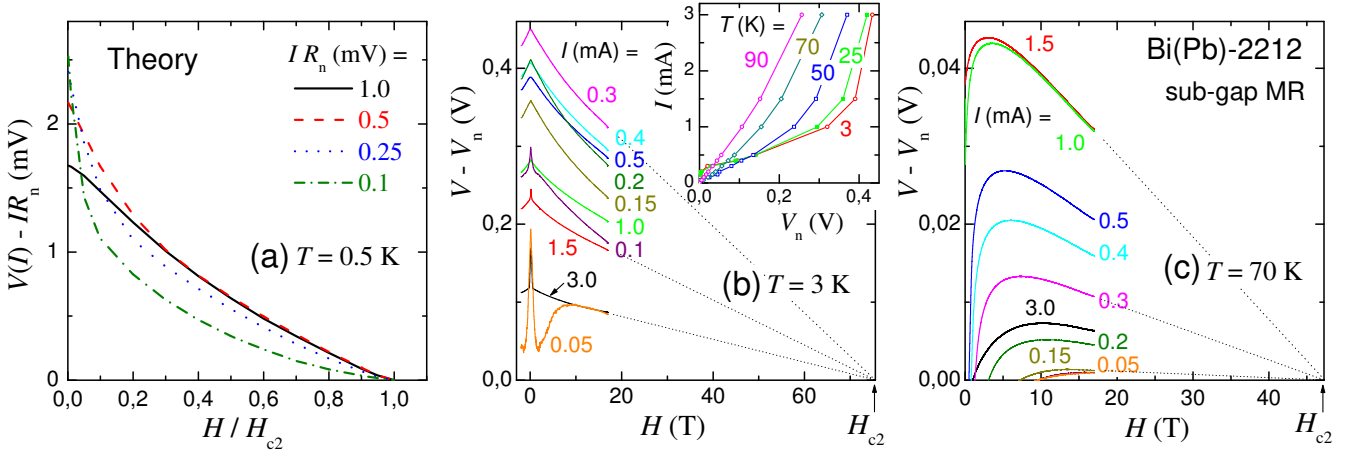}
\caption{\label{Fig7_V_I_H} (Color online). Magnetic field
dependence of the voltage at a given current for (a) theoretical
calculation, current levels correspond to horizontal lines in Fig.
\ref{Fig3 IV_H_ALL} (c), and for Bi(Pb)-2212 mesa at (b) $T=3$
K and (c) $T=70$ K at different bias currents. Inset in
(b) shows normal-state $I-V$s obtained by adjustment of $V-V_n$
so that they vanish at the same $H=H_{c2}$, as indicated by dotted
lines in (b) and (c). }
\end{figure*}

\subsection{Scaling of the sum-gap peak}

As seen from Fig. \ref{Fig4_dIdVvsH_ALL} (c), at low $T$
$dI/dV(V)$ characteristics of the Bi(Pb)-2212 mesa, normalized by
the normal state $dI/dV$, behaves in a conventional manner: the
sum-gap peak is smeared and moves to lower voltage with increasing
the $c-$axis component of the field. Solid squares in Fig.
\ref{Fig6_IVH_AR} (a) represent the corresponding peak resistance
as a function of field. Apparently, it follows the same tendency
of quasi-linear grows as for Nb-junction, shown in Fig.
\ref{Fig5_NbBCSSum} (c). This allows an approximate estimation of
$H_{c2}$ by linear extrapolation of the curves to unity, as shown
by dotted lines.

Figure \ref{Fig6_IVH_AR} (b) shows magnetic field dependence of
$dI/dV(V)$ for the mesa from Fig. \ref{Fig3 IV_H_ALL} (d) at
$T=75$ K. Here we also normalized the $dI/dV(V)$ curves
by that above $T_c$. It is seen that the general behavior of the
tunneling MR is the same as at low $T$: the sub-gap conductance
increases (negative MR) at the expense of the sum-gap conductance
peak (positive MR) in the state-conserving manner. The corresponding
peak resistances are shown by open circles in Fig. \ref{Fig6_IVH_AR} (a).
However, the sum-gap peak is no longer moving to lower voltages with increasing
$H$, but instead spreads out to higher voltages.
As discussed in Ref. \cite{KrPhC2004}, such a behavior is not
totaly unusual. As a matter of fact the upturn of the peak voltage
for high $H$ and $T$ is also observed in theory and
for Nb/AlAlOx/Nb junctions, even though in a smaller
scale, see curves at $T=4.7$ K in Fig.~\ref{Fig5_NbBCSSum} (d). As
discussed above, the outward motion of the peak is the consequence
of smearing of the peak in DOS, rather than the actual increase of
$\Delta$. This is also the reason why $V_p$ does not go to zero at
$H\rightarrow H_{c2}$, despite $\Delta$ does vanish.

Figure \ref{Fig6_IVH_AR} (c) shows the ratio of conductances at
$H=10$~T oriented parallel to $ab$-planes (which, as demonstrated in Fig.
\ref{Fig3 IV_H_ALL} (d) is equivalent to zero-field) and in the $c$-axis direction. Such a normalization
perfectly removes all field independent features and
allows observation with unprecedented clarity superconducting parts of the spectra
at $T\rightarrow T_c$. In Fig.~\ref{Fig6_IVH_AR} (c) we clearly see the sum-gap peak at $T=85$ K,
which is practically indistinguishable in $dI/dV(V)$ characteristics.
Remarkably, we can observe the peak even at $T=95$~K (see the inset),
which is above the critical temperature for appearance of phase-coherence in the $c$-axis direction
$T_c^{phase} =91-92$ K. Apparently the superconducting gap
is still present at 95 K, but its value is rapidly decreasing at
this temperature, as it does in BCS theory close to the mean-field
$T_{c0}$, see Fig.~\ref{Fig2_IVvsT_H0_All} (f). This is consistent
with the conclusion of Ref. \cite{SecondOrder} that
superconductivity in near optimally doped Bi-2212 appears by means
of the second-order phase transition in the conventional
BCS-manner. However, the true thermodynamic mean-field critical
temperature $T_{c0} \simeq 96$ K is some-what higher than
$T_c^{phase}$.

\subsection{Scaling of the sub-gap voltage}

As seen from Fig. \ref{Fig3 IV_H_ALL} (c), the voltage at
a given current is decreasing with increasing field and reaches
the normal state value $V_n(I)=I R_n$ at $H=H_{c2}$.
Fig.~\ref{Fig7_V_I_H}~(a) shows the corresponding values of $V(I,H)$
for four bias currents, indicated by horizontal lines in Fig.
\ref{Fig3 IV_H_ALL} (c), as a function of $H/H_{c2}$. It is seen
that in a wide bias range (an order of magnitude) $V(I,H)-IR_n$
have the same magnitude and disappears in a linear manner at
$H\rightarrow H_{c2}$. This fairly good and almost
bias-independent scaling is the consequence of the almost parallel
shift of $I-V$s (in the semi-logarithmic scale) with increasing $H$, as shown in Figs.~\ref{Fig3
IV_H_ALL}~(c), . The $dI/dV(V)$
curves behave in the same manner, as shown in Fig.
\ref{Fig4_dIdVvsH_ALL}.

Fig.~\ref{Fig3 IV_H_ALL}~(d) demonstrates that for Bi(Pb)-2212
such parallel translation of $I-V$ curves is even more impressive and
expands to almost two orders of magnitude in bias current. However, there is one major obstacle for
determination of the $H_{c2}$ from such the scaling. Namely, the
shape of the $I-V$ in the normal state, i.e., with completely
suppressed superconductivity but at low $T<T_c$, is unknown. We
can only say for sure that it remains non-linear above $T_c$ as a
result of thermal-activation $c$-axis transport (not necessarily
connected with the pseudogap) \cite{Katterwe2008}. To go around
this problem, we note that according to Fig.~\ref{Fig7_V_I_H} (a),
voltages at all bias levels must reach the normal state values
$V_n(I)$ in a quasi-linear manner at the same $H=H_{c2}$.
Therefore, we used $V_n(I)$ as an adjustable parameter for each
bias current, so that all $V(I)-V_n(I)$ voltages go to zero at the
same point $H=H_{c2}(T)$, as indicated by dotted lines in Figs.
\ref{Fig7_V_I_H} (b) and (c). Thus obtained normal state $I-V$
curves at $T<T_c$ are shown in the inset of Fig. \ref{Fig7_V_I_H}
(b).

Figure \ref{Fig7_V_I_H} (b) and (c) show
$H_{\perp}$-dependence of the sub-gap voltages at different bias
currents, for (b) $T=3$ K and (c) 70 K. We observe
a quasi-linear reduction of voltages, consistent with theoretical
curves in Fig. $\ref{Fig7_V_I_H}$ (a).
The extrapolated value of $H_{c2}$ at $T=3$ K is $75.5 \pm 14.5$T.
A significant uncertainty is due to the need for remote
extrapolation from the maximum available field of 17 T. However,
at low $T$ it is not crucially affected by the adjustment of
$V_n$. Indeed, already from the raw data in Fig. \ref{Fig3
IV_H_ALL} (d) it is seen that in fields from 0 to 17 T
the $I-V$ curves made approximately $1/4 - 1/5$ of the journey to the
normal state $I-V$ at $T=90$K. Assuming a linear dependence of
$V(H)$, this provides a similar estimation of $H_{c2} \sim 68-85$
T.

\begin{figure*}
\includegraphics[width=7.0in]{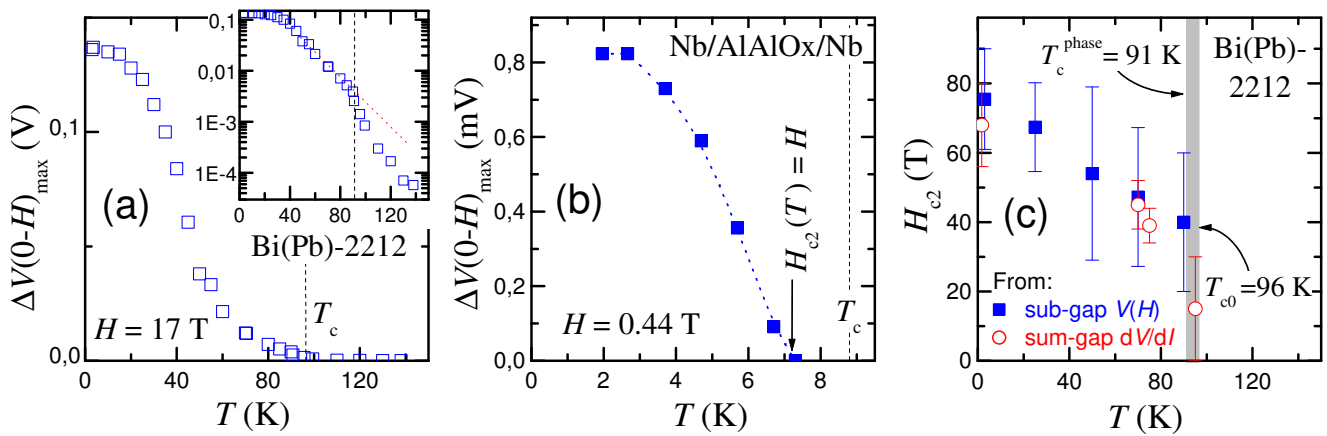}
\caption{\label{Fig8_Hc2HBi2212} (Color online). Temperature
dependence of the maximum magnetoresistance at a constant current, $\Delta V(0-H)=V(I,H=0)-V(I,H)$, (a) for Bi(Pb)-2212
mesas at $H=17$ T, (b) for the
Nb/AlAlOx/Nb junction at $H \simeq 0.44$~T with the same $H/H_{c2}(0)$ ratio. The inset in (a) shows the same data
in the semi-logarithmic scale. It is seen that the negative
sub-gap magnetoresistance decays almost exponentially with
increasing $T$, but retains a finite value at $T>T_c$. (c)
Extracted upper critical field in the $c$-axis direction for
Bi(Pb)-2212.}
\end{figure*}

\subsection{Temperature dependence of the maximum $c-$axis magnetoresistance}

Fig. \ref{Fig8_Hc2HBi2212} (a) shows temperature dependence of the
maximum measured shift of voltage at a constant current upon
variation of the $c$-axis magnetic field from 0 to 17 T. It is
seen that the negative sub-gap tunneling MR is rapidly decreasing
with increasing temperature, cf. scales in Figs. \ref{Fig7_V_I_H}
(b) and (c). This makes determination of $H_{c2}$ at elevated $T$ less confident
because it becomes more sensitive to exact values of $V_n$. For
comparison, Fig. \ref{Fig8_Hc2HBi2212} (b) shows similar data for
the Nb-junction at $H=0.44$ T, which corresponds to the
same ratio $H/H_{c2}(T=0) \simeq 0.24$ as for Bi(Pb)-2212 from
panel (a). The tunneling MR in the Nb-junction vanishes at
$T\simeq 7$~K$<T_c$, at which $H_{c2}(T) =H$, see Fig.
\ref{Fig5_NbBCSSum} (e).

From comparison of Figs. \ref{Fig8_Hc2HBi2212} (a) and (b) it is
seen that behavior of the maximum MR is qualitatively similar
for Nb and Bi(Pb)-2212. However, for Bi(Pb)-2212 it
becomes more smeared and fuzzy at $T\rightarrow T_c$. To some
extent the smoother $T-$dependence for Bi-2212 may be
caused by the $d$-wave symmetry, because nodal QP's are
more prone to thermal activation above the gap at elevated $T$.
Still this does not explain everything.

The inset in Fig. \ref{Fig8_Hc2HBi2212} (a) shows the same data
for Bi(Pb)-2212 in the semi-logarithmic scale. It is seen that the
sub-gap MR start to rapidly decrease (approximately exponentially)
with increasing temperature at $40$~K$< T < T_c$ (as indicated by
the dotted line), experiences an additional drop at $T_c$ (marked
by the dashed vertical line) and then continued to decrease at
even faster rate above $T_c$. We assume that the remaining small
negative MR above $T_c$ is a consequence of fluctuation
superconductivity, which becomes progressively less sensitive to
magnetic field due to growth of the effective $H_{c2}^* \propto
T-T_c$ above $T_c$ \cite{Varlamov}.

\subsection{Extraction of $H_{c2}$ for Bi-2212}

As follows from Fig. \ref{Fig5_NbBCSSum} (c), scaling of
sum-gap magnetoresistance provides the most accurate way of
determination of $H_{c2}$. The corresponding data for Bi(Pb)-2212 are shown in Fig. \ref{Fig6_IVH_AR} (a).
Data for 1.7 K and 75 K correspond to Figs. \ref{Fig4_dIdVvsH_ALL} (d) and
\ref{Fig6_IVH_AR} (b). The upper critical field
is estimated using a linear extrapolation towards the normal
resistance, as indicated by dotted lines. There is a
significant uncertainty in such extrapolation, however it is more
robust than that made from sub-gap MR, see Fig. \ref{Fig7_V_I_H},
because the normal resistance at a large bias is more unambiguous
and has a weak $T$-dependence
\cite{KrasnovPRL2000,Katterwe2008}, as seen from
Fig.~\ref{Fig2_IVvsT_H0_All} (c).

Figure \ref{Fig8_Hc2HBi2212} (c) shows the extracted $T$-dependence of the upper critical field for the slightly
overdoped Bi(Pb)-2212 single crystal. Solid squares and open
symbols are obtained from analysis of scaling of the sub-gap
voltage, Figs. \ref{Fig7_V_I_H} (b) and (c), and the sum-gap resistance,
Fig. \ref{Fig6_IVH_AR} (a), respectively. $H_{c2} (T=0)$ is $\sim
70$ T and certainly decreases with increasing $T$. As mentioned in
the introduction, one of the principle questions is
whether $H_{c2}$ goes to zero at $T_c$. This appears to be a
difficult question. First, the notion of $T_c$ is fuzzy, as
indicated by the gray area in Fig. \ref{Fig8_Hc2HBi2212}. We see
the clear presence of the superconducting gap in the DOS above the
phase coherent $T_c^{phase}\simeq 91$ K up to the mean-field
$T_{c0}\simeq 96$ K \cite{SecondOrder}. The main experimental
challenge is associated with a very small sub-gap MR close to
$T_c$, see Fig. \ref{Fig8_Hc2HBi2212} (a). However, at $T=95$ K
any sign of the sum-gap peak is absent at $H=15$ T, which provides
a rough estimation for the last point in the $H_{c2}(T)$ diagram
of Fig. \ref{Fig8_Hc2HBi2212} (c).

\section{Conclusions}

We performed a detailed comparative analysis of tunneling
magnetoresistances in conventional low-$T_c$ Nb/AlAlOx/Nb
junctions, small Bi(Pb)-2212 intrinsic Josephson junctions and
microscopic calculations. It
was found that magnetotunneling in slightly overdoped Bi(Pb)-2212
is qualitatively similar to that in conventional BCS-type
superconductors.

From the data presented above it is clearly seen that both
temperature and magnetic field suppress superconductivity in
Bi(Pb)-2212 in the state conserving manner. Magnetotunneling
provides a particularly clear demonstration of this: due to
conservation of states the MR changes sign from the negative in the
sub-gap, to the positive at the sum-gap bias.
Continuing strong MR well above the sum-gap peak
with powers up to several times that on the peak indicate that the
mesa remains in the superconducting state and the peak is not
caused by self-heating. This clearly demonstrates that intrinsic tunneling can provide unambiguous
information about bulk electronic spectra of Bi-2212.

Observation of state conservation implies that QP states released
upon suppression of superconductivity by magnetic field are not
taken over by a competing order, like charge or spin density wave.
In other words, there is no field-induced non-superconducting
order in slightly overdoped Bi(Pb)-2212. The situation may,
however be different for underdoped Bi-2212, for which
non-state-conserving characteristics have been reported
\cite{SuzukiMR,Gupta}.

We derived theoretically and verified experimentally scaling laws
of various magnetotunneling parameters. Those scaling laws were
employed for accurate extraction of the upper critical fields and
in the case of Nb provided a clear evidence for the existence of
an extended region of surface superconductivity at
$H_{c2}<H<H_{c3}$.

For Bi(Pb)-2212, it was found that $H_{c2}(T=0) \simeq 70$~T and
decreases significantly upon approaching $T_c$. The parameters of
Bi(Pb)-2212 were obtained from self-consistent analysis of
magnetotunneling data at different levels of bias, dissipation
powers and different mesa sizes, which precludes the influence of
self-heating. The amplitude of the sub-gap magnetoresistance is
suppressed exponentially at $T>T_c/2$. It remains negative,
although very small, above $T_c$, probably indicating existence of
an extended fluctuation region.

We conclude in general that intrinsic magnetotunneling in small
mesa structures is a very powerful tool for analysis of bulk
superconducting properties of cuprates.

\begin{acknowledgments}
We are grateful to A. Golubov for assistance with numerical
simulations, D. Haviland for providing Nb/AlAlOx/Nb junctions, and to
A. Kordyuk, S. Borisenko, D. Munzar and Ch. Bernhard for stimulating discussions.
Financial support from the Swedish Research Council, the SU-Core
Facility in Nanotechnology and the K.{\&}A. Wallenberg foundation
is gratefully acknowledged.
\end{acknowledgments}

\end{document}